**Emergence of Phonon Hydrodynamics as a Non-Hermitian Symmetry-Breaking Phase Transition**

Xin Qian[1], Guanda Quan[1], Ning Wang [2,*], Yaoming Chu[2], Run Hu[1], Te-Huan Liu[1], Xiangfan Xu[3], Jianming Cai[2,†], and Ronggui Yang[1,4,‡]

[1]School of Energy and Power Engineering, Huazhong University of Science and Technology, Wuhan 430074, China

[2] School of Physics, International Joint Laboratory on Quantum Sensing and Quantum Metrology, Hubei Key Laboratory of Gravitation and Quantum Physics, Institute for Quantum Science and Engineering, Center for Intelligence and Quantum Science (CIQS), Huazhong University of Science and Technology, Wuhan 430074, China

[3]Center for Phononics and Thermal Energy Science, School of Physics Science and Engineering, Tongji University, Shanghai 200092, China

[4]School of Mechanics and Engineering Science, Peking University, Beijing, 100871, China

*Contact author: ningwang@hust.edu.cn

†Contact author: jianmingcai@hust.edu.cn

‡Contact author: ronggui@pku.edu.cn

## Abstract

Phonon hydrodynamics enables collective heat flow beyond the classical diffusive limit and offers a promising route to controlling thermal transport in low-dimensional materials and nanoscale devices. However, quantitative criteria for determining when this hydrodynamic state emerges and where its boundaries lie remain a longstanding challenge in condensed matter physics. Here, we show that the onset of phonon hydrodynamics can be formulated as a non-Hermitian dynamical transition that converts microscopic phonon scattering properties into predictive boundaries for wavelike heat transport. Starting from the Peierls–Boltzmann transport equation, we derive an effective Schrődinger equation for the coupled dynamics of temperature and phonon drift fields, allowing the hydrodynamic regime to be identified directly from material transport properties. Within this framework, a pair of exceptional points quantitatively define the lower and upper boundaries of the hydrodynamic window, providing a critical design criterion beyond the qualitative scattering-rate inequalities proposed by Guyer and Krumhansl six decades ago. Furthermore, we derive an analytical criterion for the existence of hydrodynamic transport, showing that a finite hydrodynamic window can be supported only when the effective thermal diffusivity exceeds the phononic momentum diffusivity. Using *ab initio* phonon properties as inputs, our theory accurately captures the experimentally observed temperature- and length-dependent ultrafast second-sound dynamics. These results establish a symmetry-based framework for understanding collective energy transport and provide quantitative design principles for identifying the materials, temperatures, and characteristic length scales that support wavelike heat transport in low-dimensional materials and functional devices.

*Introduction* – Spontaneous symmetry breaking (SSB) is a fundamental concept governing a wide range of phenomena [1], from the Higgs mechanism [2] in particle physics to phase transitions in condensed matter. For example, superconductivity originates from the breaking of U(1) symmetry [3,4]. Ferroelectricity arises as a result of the spontaneous breaking of the lattice centrosymmetry [5], and charge density waves are underpinned by the breaking of periodic translational symmetry [6]. The engineering of SSB has enabled locating phase boundaries, classifying emergent states, and ultimately controlling transitions for developing functional devices [7-9].

This symmetry-based perspective has increasingly been extended to open and nonequilibrium dynamical systems at the nanoscale through non-Hermitian physics. By mapping the governing dynamical equation of the system to a formal Schrödinger equation $i\frac{\partial}{\partial t}|\psi\rangle = \mathcal{H}|\psi\rangle$, an effective Hamiltonian $\mathcal{H}$ can be extracted, enabling identification of exceptional points (EPs) where symmetry breaking occurs [10,11]. In contrast to isolated quantum systems, these effective Hamiltonians are generally non-Hermitian (*i.e.* $\mathcal{H}^\dagger \neq \mathcal{H}$) due to the existence of losses or gains. For example, $\mathcal{H}$ of a wave system typically obey parity-time (PT) symmetry. When the system is tuned across the EP, PT symmetry breaking leads to captivating phenomena, including localization transitions [12], single-mode lasing [13], topological and non-reciprocal energy transfer [14], and others [15-17]. Recently, macroscopic heat transfer systems have become a fruitful playground for exploring non-Hermitian physics [18]. For example, a pair of counter-rotating rings with convective coupling governed by the conduction-convection equation can be mapped to an effective SE with APT [19] or PT symmetry [20], allowing precise control of the propagating direction of the temperature fields. This raises a broader possibility: can symmetry breaking similarly organize the transport regimes that emerge collectively from microscopic many-body dynamics?

Collective phonon transport provides a particularly compelling setting in which to address this question. While thermal energy propagates diffusively in bulk materials, the transport of energy carriers, primarily phonons in semiconductors and dielectrics, can strongly deviate from the diffusive random-walk picture[21] in nanostructures or near localized hotspots when the characteristic length becomes comparable to phonon mean free paths (MFPs). Beyond the extensively studied diffusive and quasi-ballistic regimes, phonon hydrodynamics emerges as a collective transport regime in which the momentum-conserving normal scattering processes dominate over the momentum-dissipating processes such as Umklapp, defect, isotope, and boundary scattering events [22-24]. Under steady-state temperature gradients, the heat flux across a nano-strip can develop a parabolic profile resembling the Poiseuille flow in fluid tubes, whereas transient thermal perturbations can propagate in a wavelike manner, known as second sound [25-28]. Recent experimental observations of second sound at elevated temperatures [22,29,30] have stimulated intensive interest in phonon hydrodynamics, pointing to opportunities for manipulating heat flow in micro/nano-structures by exploiting the fluid-like motion of collective phonons and for rapid wavelike heat dissipation in electronics [31].

Despite these advances, quantitative criteria for phonon hydrodynamics remain lacking, leaving its viable window in terms of material, temperature, and length scale unpredictable without case-by-case experiments or costly first-principles computations. Phonon hydrodynamics is currently understood as the competition between momentum-conserving and momentum-dissipating scattering. The celebrated criterion for achieving hydrodynamic transport, proposed by Guyer and Krumhansl [28] six decades ago, qualitatively defined the hydrodynamic regime: phonon hydrodynamics occurs when $\gamma_U \ll v_g/l \ll \gamma_N$, where $\gamma_U$ and $\gamma_N$ are scattering rates of Umklapp and normal processes, $v_g$ denotes the average phonon group velocity, and $l$ is the characteristic system size. Although the development of *ab initio* Peierls-Boltzmann transport equation (PBTE) enabled parameter-free prediction of lattice thermal conductivity[32-34] and the modeling of spatiotemporal temperature evolutions at nanoscales [35-37], these theoretical models do not provide a quantitative criterion for identifying the critical boundaries separating the quasiballistic, diffusive, and hydrodynamic phonon transport regimes.

In this work, we establish a non-Hermitian framework for collective phonon transport and uncover the spontaneous symmetry breaking mechanisms in phonon hydrodynamics. From the PBTE, we derive an effective Schrödinger equation (SE) that governs the coupled dynamics of temperature and collective phonon drift fields when the heating frequency is much slower than the phonon scattering rates. The Hamiltonian obtained from the effective SE is non-Hermitian and exhibits the anti-parity-time (APT) symmetry. Within this framework, spontaneous APT symmetry breaking gives rise to hydrodynamic second sound in a phase confined by a pair of exceptional points (EPs). The EPs define computable lower and upper thresholds distinguishing overdamped relaxation from wave-like second sound. Furthermore, we derive an analytical criterion for identifying materials that can support hydrodynamic transport, revealing that a finite hydrodynamic window exists only when the thermal diffusivity exceeds the phononic momentum diffusivity. Using *ab initio* phonon properties, we accurately reproduce the second sound dynamics measured by transient grating experiments at different temperatures and length scales. Unlike previous hydrodynamic formulations, the present framework directly maps microscopic phonon properties into computable hydrodynamic boundaries and experimentally verifiable phase diagrams. These findings offer quantitative design guidelines to identify the materials, temperatures, and characteristic length scales that are capable of supporting wavelike heat transport in low-dimensional materials and functional devices.

*Effective Schrödinger equation of thermal waves* – In this section, we demonstrate that the mode-resolved PBTE can be projected onto two collective variables—the local temperature rise and phonon drift velocity—yielding an effective two-level non-Hermitian description of thermal waves. As shown in Fig. 1(a), we account for the collective incoherent transport of phonon quasiparticles to formulate the non-Hermitian theory. Microscopic phonon dynamics under a temperature gradient is described by the PBTE:

$$\frac{\partial N_\mu}{\partial t} + \boldsymbol{v}_\mu \cdot \boldsymbol{\nabla} N_\mu = -\frac{N_\mu - N_\mu^0(T)}{\tau_\mu^R} - \frac{N_\mu - N_\mu^D(T, \boldsymbol{u})}{\tau_\mu^N} \quad (1),$$

where $T$ is the local temperature, $N_\mu$ is the phonon population of the mode $\mu = (\boldsymbol{q}_\mu, s)$ with wavevector $\boldsymbol{q}_\mu$ and polarization $s$, and $\boldsymbol{v}_\mu$ is the phonon group velocity. $N_\mu^0(T) = N_\mu^0(T_0) + C_\mu \Delta T/(\hbar\omega_\mu)$ is the local equilibrium distribution, with $N_\mu^0$ denoting the Bose-Einstein distribution and $\Delta T = T - T_0$ being the local temperature rise from the equilibrium $T_0$; $C_\mu$ and $\omega_\mu$ are the specific heat and frequency of the phonon mode $\mu$, respectively; and $N_\mu^D(T, \boldsymbol{u}) = N_\mu^0(T_0) + C_\mu \Delta T/(\hbar\omega_\mu) + C_\mu T_0 \boldsymbol{q}_\mu \cdot \boldsymbol{u} \,/\, (\hbar\omega_\mu^2)$ is the drifted local equilibrium [38], with $\boldsymbol{u}$ the collective drift velocity of phonons. $\tau_\mu^N$ and $\tau_\mu^R$ are the relaxation time of momentum-conserving normal (N) scattering events and momentum-destroying (Umklapp and isotope) resistive (R) scattering events, respectively. Using the Fourier transform ($t \to i\omega, \boldsymbol{\nabla} \to -i\boldsymbol{k}$), the nonequilibrium deviation in population $n_\mu = N_\mu - N_\mu^0(T_0)$ can be solved as:

$$n_\mu = \chi_\mu \left( \frac{C_\mu \Delta T}{\hbar\omega_\mu} + \eta_\mu \frac{C_\mu T_0}{\hbar\omega_\mu^2} \boldsymbol{q}_\mu \cdot \boldsymbol{u} \right) \tag{2}$$

where $\eta_\mu = \tau_\mu \,/\, \tau_\mu^N$ is the relative strength of N scattering events; $\chi_\mu = \left(1 + i\omega\tau_\mu - i\boldsymbol{k} \cdot \boldsymbol{F}_\mu\right)^{-1}$ is the linear susceptibility; $\omega$ and $\boldsymbol{k}$ are the frequency and wavevector of thermal oscillations; $\boldsymbol{F}_\mu = \boldsymbol{v}_\mu \tau_\mu$ is the phonon mean free displacement.

By applying the constraints of momentum conservation of N scattering events and energy conservation, we can derive a set of linear equations from which the $\Delta T$ and $\boldsymbol{u}$ can be solved:

$$\begin{bmatrix} \mathbb{A}_{TT} & T_0 \mathbb{A}_{cp}^T \\ \mathbb{A}_{cp} & \mathbb{A}_{uu} \end{bmatrix} \cdot \begin{bmatrix} \Delta T \\ \boldsymbol{u} \end{bmatrix} = \boldsymbol{0} \tag{3}$$

The matrix elements are expressed as a summation over phonon modes across the first Brillouin zone:

$$\mathbb{A}_{TT} = \sum_\mu \frac{C_\mu}{\tau_\mu} (1 - \chi_\mu) \tag{4}$$

$$\mathbb{A}_{cp} = -\sum_\mu \frac{C_\mu \boldsymbol{q}_\mu}{\omega_\mu \tau_\mu^N} \chi_\mu \tag{5}$$

$$\mathbb{A}_{uu} = T_0 \sum_\mu \frac{C_\mu \boldsymbol{q}_\mu \boldsymbol{q}_\mu}{\omega_\mu^2 \tau_\mu^N} (1 - \eta_\mu \chi_\mu) \tag{6}$$

See Note 1 of Supplementary Materials for detailed derivations. The direct numerical solution of Eq. (3) has been implemented in our previous work on developing the API Phonons package [37,39]. Through Eqs. (3-6), the dynamics of a large number of phonon modes are reduced to a set of coupled equations governing local equilibrium fields. The microscopic mode dependence is preserved via effective transport coefficients, which take the form of summations over mode-resolved phonon properties.

We focus on collective thermal responses with timescales longer than individual phonon relaxation times, for which phonons undergo many scattering events during one thermal oscillation period, such that $\omega\tau_\mu \ll 1$. This slow-varying approximation is valid, since pump-probe thermal measurements are within a few nanoseconds, while the phonon lifetimes are on the order of picoseconds. Matrix elements in Eq. (3) can then be simplified as:

$$\mathbb{A}_{TT} \approx i\omega\tilde{C} + k^\alpha \tilde{\kappa}^{\alpha\beta} k^\beta \tag{7}$$

$$\mathbb{A}_{cp}^\alpha \approx -\tilde{\Pi}^{\alpha\beta} i k^\beta \tag{8}$$

$$\mathbb{A}_{uu}^{\alpha\beta} \approx T_0\left(\tilde{\Gamma}^{\alpha\beta} + \tilde{A}^{\alpha\beta} i\omega + \tilde{\mathrm{M}}^{\alpha\beta\xi\zeta} k^{\xi} k^{\zeta}\right) \tag{9}$$

where the subscripts $(\alpha, \beta, \xi, \zeta)$ denote the Cartesian coordinates, and the Einstein summation convention is used. See Note 2 of Supplementary Materials for detailed derivations. $\tilde{C}$, $\tilde{\kappa}^{\alpha\beta}$, $\tilde{\Pi}^{\alpha\beta}$, $\tilde{\Gamma}^{\alpha\beta}$, $\tilde{A}^{\alpha\beta}$ and $\tilde{\mathrm{M}}^{\alpha\beta\xi\zeta}$ are the effective values of specific heat, thermal conductivity, momentum transport coefficient, momentum dissipation rate, specific momentum intensity, and phonon viscosity, respectively, whose expressions are given in Table I. All coefficients in Eqs. (7-9) are denoted by a tilde (~), indicating they are corrected by a mode-resolved suppression function $S_\mu(\boldsymbol{k}) = 1/\left[1 + \left(\boldsymbol{k} \cdot \boldsymbol{F}_\mu\right)^2\right]$, which is identical to the formulation derived by Minnich [40]. The suppression function accounts for the departure of the phonon distribution from the bulk limit when the characteristic thermal length approaches the mode-resolved phonon mean free displacement.

**Table I. Effective transport properties**

| Physical quantities | Expression | Physical quantities | Expression |
|---|---|---|---|
| Effective specific heat | $\tilde{C} = \sum_\mu C_\mu S_\mu$ | Effective momentum dissipation rate | $\tilde{\Gamma}^{\alpha\beta} = \sum_\mu \frac{\eta_\mu}{\tau_\mu^R} \frac{C_\mu q_\mu^\alpha q_\mu^\beta}{\omega_\mu^2} S_\mu$ |
| Effective thermal conductivity | $\tilde{\kappa}^{\alpha\beta} = \sum_\mu C_\mu v_\mu^\alpha F_\mu^\beta S_\mu$ | Specific momentum intensity | $\tilde{A}^{\alpha\beta} = \sum_\mu \frac{C_\mu q_\mu^\alpha q_\mu^\beta}{\omega_\mu^2} \eta_\mu (2 - \eta_\mu) S_\mu$ |
| Effective momentum transport coefficient | $\tilde{\Pi}^{\alpha\beta} = \sum_\mu \eta_\mu \frac{C_\mu q_\mu^\alpha v_\mu^\beta}{\omega_\mu} S_\mu$ | Effective phonon viscosity | $\tilde{\mathrm{M}}^{\alpha\beta\xi\zeta} = \sum_\mu \eta_\mu \frac{C_\mu q_\mu^\alpha q_\mu^\beta v_\mu^\xi F_\mu^\zeta}{\omega_\mu^2} S_\mu$ |
| Suppression function $S_\mu(\boldsymbol{k}) = 1/\left[1 + \left(\boldsymbol{k} \cdot \boldsymbol{F}_\mu\right)^2\right]$ | | | |

Substituting Eqs. (7-9) into Eq. (3), and performing the inverse Fourier transform to $\omega$, we can derive an effective SE of thermal wave mechanics in the space of thermal wavevector $\boldsymbol{k}$. In the one-dimensional (1D) case, the effective SE is written as:

$$i \frac{\partial}{\partial t} |\psi\rangle = \mathcal{H}(k) |\psi\rangle \tag{10}$$

$|\psi\rangle = \left[\sqrt{\tilde{C}}\theta,\ \sqrt{\tilde{A}^{xx}}\, u^x\right]^T$ is the state vector, and $\theta = \Delta T / T_0$ is the normalized temperature rise. Note that $|\psi\rangle$ here is not quantum mechanical, but a state vector containing local equilibrium fields that are classical. $\mathcal{H}(k)$ is the effective Hamiltonian of thermal waves:

$$\mathcal{H}(k)=\begin{bmatrix}-i\tilde{\alpha}k^2 & \tilde{v}_0 k\\ \tilde{v}_0 k & -i(\tilde{\eta}k^2+\tilde{\gamma})\end{bmatrix} \tag{11}$$

$k$ represents the wavevector of the thermal fields ($\theta$ and $\boldsymbol{u}$). $\tilde{\alpha}=\tilde{\kappa}^{xx}/\tilde{C}$ is the effective thermal diffusivity; $\tilde{\eta}=\widetilde{\mathrm{M}}^{xxxx}/\tilde{A}^{xx}$ is the momentum diffusivity; $\tilde{\gamma}=\tilde{\Gamma}^{xx}/\tilde{A}^{xx}$ is the momentum dissipation coefficient, and $\tilde{v}_0=\left|\widetilde{\Pi}^{xx}\right|/\sqrt{\tilde{A}^{xx}\tilde{C}}$ is the inviscid phase velocity of thermal waves, *i.e.,* phonon viscosity is neglected. At the N scattering limit ($\eta_\mu\to 1$), $\tilde{v}_0$ recovers the expression derived by Hardy and Ding [41,42]. See Notes 2-3 of Supplementary Materials for derivation of $\mathcal{H}$ in both one-dimensional and multiple-dimensional cases, respectively.

$\mathcal{H}(k)$ of collective phonon thermal wave transport obeys the APT symmetry, *i.e.* $(\mathcal{PT})^{-1}\mathcal{H}(\mathcal{PT})=-\mathcal{H}$, with $\mathcal{P}$ the parity inversion operator reversing the sign of $k$ and $u$, and $\mathcal{T}$ the time-reversal operator inverting the direction of time. APT symmetry imposes constraints on the distribution of eigen-frequencies in the complex plane and dictates the permissible thermal transport behaviors. The eigen-frequencies take two forms: purely imaginary numbers when the corresponding eigenvector satifies APT symmetric, i.e., $\mathcal{PT}|\psi\rangle=-|\psi\rangle$; or anti-conjugate pairs features opposite real parts but identical imaginary parts when the APT symmetry is broken for the eigenvector, i.e., $\mathcal{PT}|\psi\rangle\neq-|\psi\rangle$ [19]. As we shall see later, wave-like energy transport is only supported in the APT-symmetry-broken phase, where eigen-frequencies have nonzero real parts.

$\mathcal{H}(k)$ corresponds to a dissipative two-level system, as illustrated in Fig. 1(c). The imaginary diagonal elements account for the thermal decay ($-i\alpha k^2$) and the momentum decay ($-i(\tilde{\eta}k^2+\tilde{\gamma})$). The symmetric real-valued off-diagonal term $\tilde{v}_0 k$ characterizes the coupling between the temperature field and the drifting velocity. In heat conduction, each phonon mode carries both energy $\hbar\omega_\mu$ and the crystal momentum $\hbar\boldsymbol{q}_\mu$. When N scattering dominates, the crystal momentum cannot be effectively dissipated by phonon-phonon interaction. Consequently, the collective phonon drift velocity $\boldsymbol{u}$ exhibits strong dependence on the temperature field $\theta$. By contrast, when R scattering prevails ($\eta_\mu\to 0$), the momentum flux is dissipated $\widetilde{\Pi}^{xx}\to 0$, and the coupling term $\tilde{v}_0 k\propto\widetilde{\Pi}^{xx}$ vanishes.

The effective SE derived herein does not mean that the system is quantum mechanical, since the derivation originates from the semi-classical PBTE. Instead, it should be regarded as an effective dynamical equation governing the macroscopic thermal fields ($\theta$ and $\boldsymbol{u}$) emerging from microscopic phonon collisions. In deriving the effective SE, we assumed $\omega\tau_\mu \ll 1$ while imposing no restrictions on the scale of the thermal wavevector $k$. Therefore, the effective SE is valid throughout the quasiballistic, hydrodynamic, and diffusive transport regimes. Although our derivation employs a simplified Callaway scattering approximation, we show in Note 4 of the Supplementary Materials that the effective SE can recover the viscous heat equation (VHE) [43] by neglecting the nonlocal effects. Notably, the VHE was originally derived using the relaxon theory with the full scattering matrix and can be transformed back to the effective SE in the Fourier space. This shows that the APT symmetry of collective phonon transport derived in this work is robust and insensitive to the detailed formulation of the microscopic scattering matrix. As further discussed below, the simplified scattering matrix treatment still yields reasonable numerical accuracy in modeling thermal wave oscillations in ultrafast pump-probe experiments.

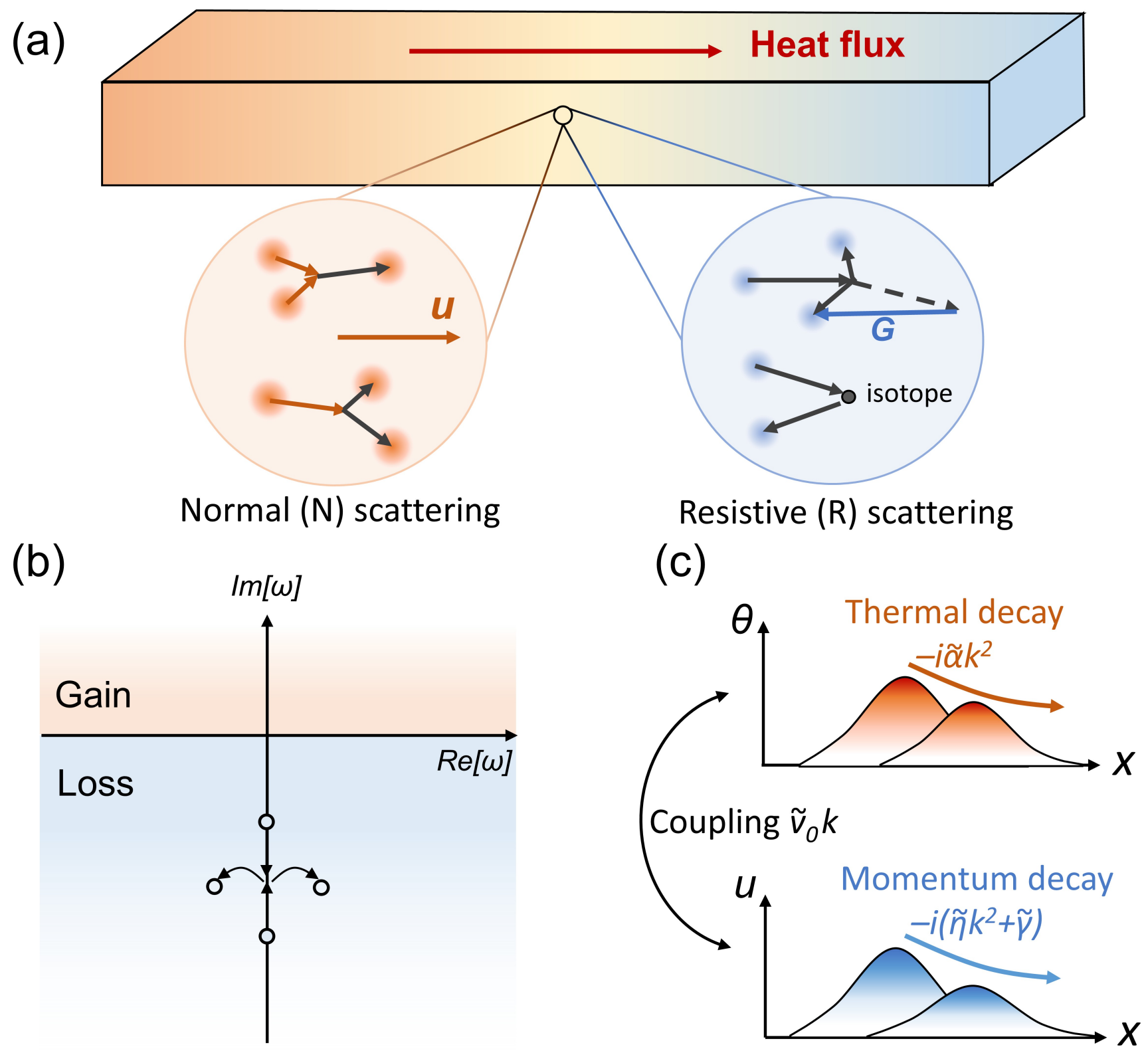

FIG. 1. Schematic of non-Hermitian physics of collective phonon transport. (a) Competition between the momentum-conserving N scattering and the momentum-destroying R scattering. $\boldsymbol{G}$ denotes a reciprocal lattice vector. (b) Eigen-frequencies of $\mathcal{H}$ possessing APT symmetry in the complex plane emerge either as a pair of purely imaginary numbers or as complex numbers with identical imaginary parts but opposite real parts. (c) Analogy of the non-Hermitian $\mathcal{H}$ to a dissipative two-level system.

*APT symmetry breaking phase transition* – Having derived the effective non-Hermitian Hamiltonian for collective phonon transport, we now investigate how its spectrum dictates the transition between overdamped thermal relaxation and wavelike second-sound propagation. Within this framework, the real part of the eigenfrequency corresponds to oscillatory thermal-wave propagation, while the imaginary part accounts for dissipation. Accordingly, the onset of second sound can be pinpointed by the emergence of nonzero real eigenfrequencies.

As demonstrated below, this transition is driven by spontaneous breaking of APT symmetry, and the associated exceptional points offer a straightforward way to identify the hydrodynamic window.

We first obtain the dispersion of supported thermal waves by solving the eigenvalue problem of $\mathcal{H}(k)$:

$$\omega_{\pm}(k) = -i\Gamma_{\theta} \pm \Omega_{\theta} \tag{12}$$

$$\Omega_{\theta} = \sqrt{(\tilde{v}_0 k)^2 - [\tilde{\gamma} - k^2(\tilde{\alpha} - \tilde{\eta})]^2/4} \tag{13}$$

$$\Gamma_{\theta} = [(\tilde{\alpha} + \tilde{\eta})k^2 + \tilde{\gamma}]/2 \tag{14}$$

where $\omega_{\pm}$ denotes the eigenfrequencies. Fig. 2 shows the dispersion of thermal waves calculated from *ab initio* phonon properties of graphite at 100 K (see Supplementary Materials Note 6 for computational details). The sign of $\Omega_{\theta}^2$ determines whether the eigenfrequencies acquire nonzero real parts. The exceptional points emerge at $\Omega_{\theta}^2 = 0$, where the two eigenmodes coalesce. While the coefficients listed in Table I depend on the wavevector through $S_{\mu}(\boldsymbol{k}) = \left[1 + \left(\boldsymbol{k} \cdot \boldsymbol{F}_{\mu}\right)^2\right]^{-1}$, $S_{\mu}$ varies much more slowly than quadratic $O(k^2)$ or quartic terms $O(k^4)$ in Eq. (13) within the range spanning across the two EP points. Hence, we can solve $\Omega_{\theta}^2 = 0$ as a polynomial equation, and a pair of EPs can be analytically located as:

$$(k_{EP1})^2 = \frac{2\tilde{v}_{ss}^2 - \sqrt{4\tilde{v}_{ss}^4 - \tilde{\gamma}^2(\tilde{\alpha} - \tilde{\eta})^2}}{(\tilde{\alpha} - \tilde{\eta})^2} \approx \left(\frac{\tilde{\gamma}}{2\tilde{v}_{ss}}\right)^2 \tag{15}$$

$$(k_{EP2})^2 = \frac{2\tilde{v}_{ss}^2 + \sqrt{4\tilde{v}_{ss}^4 - \tilde{\gamma}^2(\tilde{\alpha} - \tilde{\eta})^2}}{(\tilde{\alpha} - \tilde{\eta})^2} \tag{16}$$

Here, $\tilde{v}_{ss}^2 = \tilde{v}_0^2 + \tilde{\gamma}(\tilde{\alpha} - \tilde{\eta})/2$ is the second sound velocity with finite phonon viscosity. Fig. 2 shows the dispersion of the collective thermal waves in graphite at 100 K, in which N scattering is sufficiently strong to support phonon hydrodynamic transport. Between the two exceptional points, *i.e.* $k_{EP1} < k < k_{EP2}$, the real parts of eigenfrequencies $\mathrm{Re}[\omega_\pm]$ are nonvanishing with opposite signs; hence the $\omega_\pm$ branches correspond to forward (+) and backward (−) propagating second sound waves. Numerical results based on *ab initio* phonon property calculations show that second sound emerges on micrometer length scales in graphite with strong N scattering, consistent with TTG experiments[22,29]. This intermediate length scale arises from a physical requirement: phonons must experience sufficient momentum-conserving collisions to establish collective drift before momentum-dissipating processes induce overdamping of thermal propagation. The eigenvectors are solved as $|\psi_\pm\rangle = [\tilde{v}_0 k, \xi_\pm]$, where $\xi_\pm = i[k^2(\tilde{\alpha} - \tilde{\eta}) - \tilde{\gamma}]/2 \pm \Omega_\theta$. Within this hydrodynamic regime bounded by the EP pair, $\Omega_\theta$ is a real number, and the eigenstates cease to satisfy APT symmetry $\mathcal{PT}|\psi_\pm\rangle \neq -|\psi_\pm\rangle$. Therefore, the hydrodynamic second sound corresponds to a non-Hermitian phase transition featuring APT symmetry breaking. In this regime, the thermal-momentum coupling is sufficiently strong to induce oscillations between the temperature field and the collective drift field.

Outside the hydrodynamic window $k < k_{EP1}$ or $k > k_{EP2}$, $\mathrm{Re}[\omega_\pm]$ vanishes and the eigenmodes become fully dissipative. The two APT-symmetric regimes outside the hydrodynamic window correspond to physically distinct limits: diffusive transport at long wavelengths and quasiballistic transport at short wavelengths. Notably, the two dissipative modes $\omega_\pm$ exhibit distinct decaying rates and correspond to different dissipation mechanisms far away from the EPs. In the long-wavelength region $k \ll k_{EP1}$, the dispersion of the $\omega_+$ branch can be approximated as $\omega_+ \approx -iD_{th}k^2$, where $D_{th} = \tilde{\alpha} + \tilde{v}_0^2/\tilde{\gamma}$ denotes the apparent thermal diffusivity. The additional term $\tilde{v}_0^2/\tilde{\gamma}$ accounts for the diffusivity correction due to collective phonon transport. This indicates that even outside the hydrodynamic window, the collective effects due to strong Normal scattering can contribute to thermal transport in the bulk limit, consistent with previous studies [44,45]. At the limit of strong Umklapp scattering, $\tilde{v}_0/\tilde{\gamma}$ vanishes and $D_{th}$ reduced to the conventional kinetic thermal diffusivity $\tilde{\alpha}$. The other $\omega_-$ branch with $k \ll k_{EP1}$ governs momentum dissipation. The dispersion can be approximated as $\omega_- \approx -i\tilde{\gamma} - iD_P k^2$, where $D_P = (\tilde{\eta} - \tilde{v}_0^2/\tilde{\gamma})$ is the apparent momentum diffusivity arising jointly from viscous momentum diffusion and dissipation. The detailed derivation is presented in Note 5 of the Supplementary Materials .

In the short wavelength regime ($k \gg k_{EP2}$) outside the hydrodynamic window, thermal energy propagates in a quasiballistic manner. In this case, the diagonal elements of $\mathcal{H}$ have much larger magnitudes than the coupling element $\tilde{v}_0 k$, such that the thermal decay and the momentum decay decouple. Interestingly, the dominant dissipation channels are exchanged relative to the long-wavelength limit. By contrast, in the short-wavelength limit, thermal relaxation is governed primarily by the $\omega_- \approx -i\tilde{\alpha}k^2$ branch, whereas momentum relaxation is dominated by the $\omega_+ \approx -i(\tilde{\eta}k^2 + \tilde{\gamma})$ branch. At such small spatial scales, collective phonon drift cannot fully develop. As a result, the thermal branch reduces to the purely kinetic form $-i\tilde{\alpha}k^2$, with no contributions from the collective diffusivity $\tilde{v}_0^2/\tilde{\gamma}$. Thus, although both the long- and short-wavelength regimes are APT-symmetric and overdamped, the interchange of the dominant thermal- and momentum-relaxation channels distinguishes short-wavelength quasiballistic transport from long-wavelength diffusion.

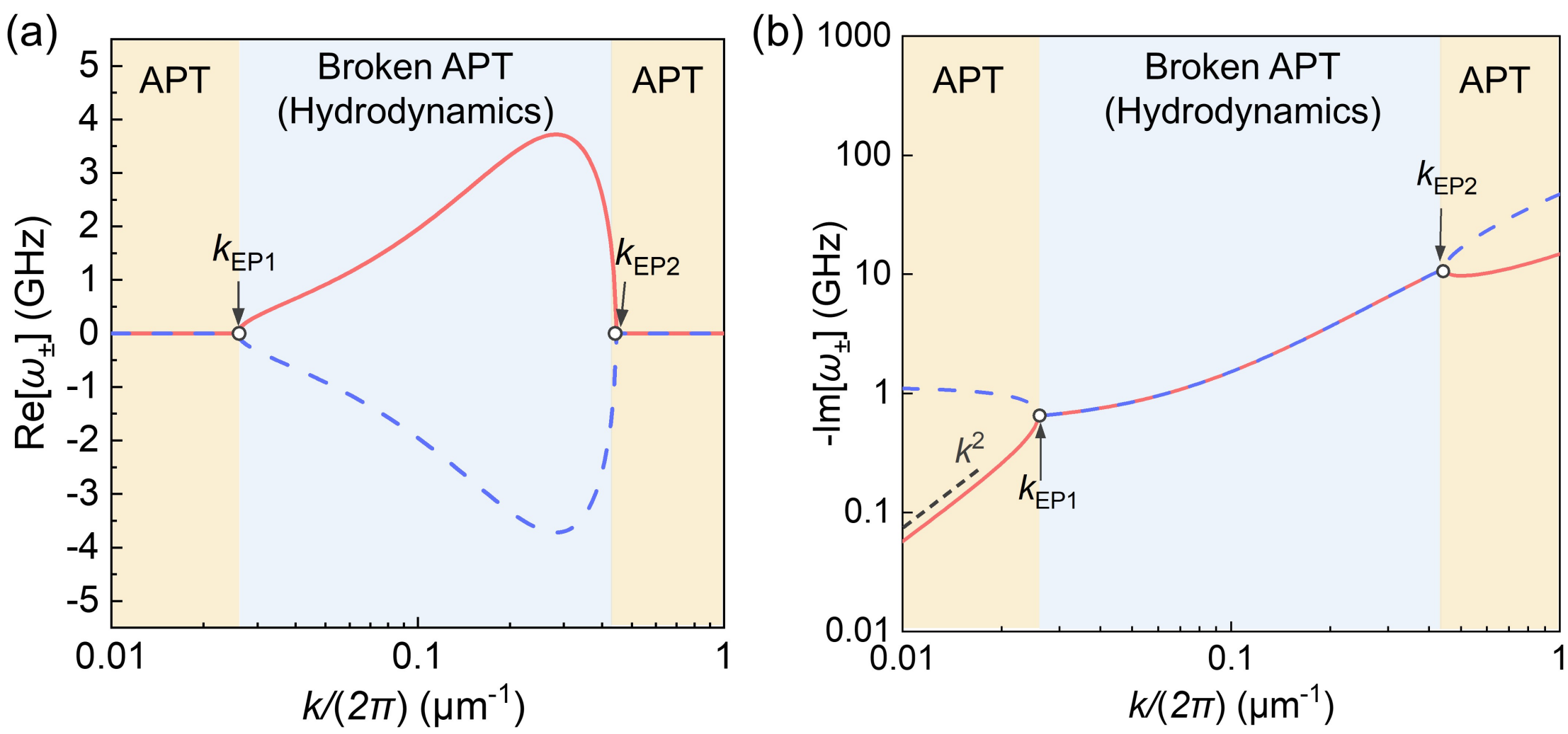

FIG. 2. Dispersions of thermal waves in graphite at 100 K. (a) Real part and (b) imaginary part of eigen-frequencies. The red solid line denotes the $\omega_+$ branch, and the blue dashed line denotes the time-reversed $\omega_-$ branch. The $k^2$-scaling rule of Fourier's law [46] is indicated by a black short-dashed line. The black circles mark the EPs calculated analytically from Eqs. (15) and (16), whose positions agree closely with the wavevectors at which the real-frequency branches bifurcate and the imaginary-frequency branches coalesce.

*Existence condition of phonon hydrodynamics* – The EP pair derived above not only delineates the boundaries of the hydrodynamic window but also provides a quantitative criterion for the existence of such a window. Requiring both boundary wavevectors, $k_{EP1}$ and $k_{EP2}$, to be real imposes the constraint $4\tilde{v}_{ss}^4 - \tilde{\gamma}^2(\tilde{\alpha} - \tilde{\eta})^2 > 0$. After simplification, this requirement yields:

$$D_{th} = \tilde{\alpha} + \tilde{v}_0^2/\tilde{\gamma} > \tilde{\eta} \tag{17}$$

Equation (17) shows that second sound can emerge over a finite range of spatial scales only when the effective thermal diffusivity, comprising both kinetic diffusion $\tilde{\alpha}$ and the collective contribution $\tilde{v}_0^2/\tilde{\gamma}$, exceeds the momentum diffusivity $\tilde{\eta}$ (Fig. 3a). Physically, when momentum perturbations decay spatially faster than temperature perturbations, the collective drift field is suppressed before sustained oscillatory coupling with the temperature field can develop. The $k_{EP1}$ and $k_{EP2}$ coalesce when the thermal diffusivity is critically matched with the momentum diffusivity, and the hydrodynamic window width becomes infinitely small, where a band touching of $\mathrm{Im}[\omega_\pm]$ occurs at the degenerate EP pair (Fig. 3b). Finally, if $D_{th} < \tilde{\eta}$, then the spontaneous APT-symmetry breaking cannot occur at any spatial scale. In this case, the thermal waves will always be overdamped (Fig 3c), and the dissipative $\mathrm{Im}[\omega_\pm]$ branches are always gapped, indicating decoupled momentum and thermal dissipation. As a result, the crossover from quasiballistic transport to diffusive conduction is continuous without any non-Hermitian phase transitions in this case. Our results show that hydrodynamic transport therefore exists only after an EP pair is created and disappears through pair annihilation when the thermal and momentum diffusivities are matched.

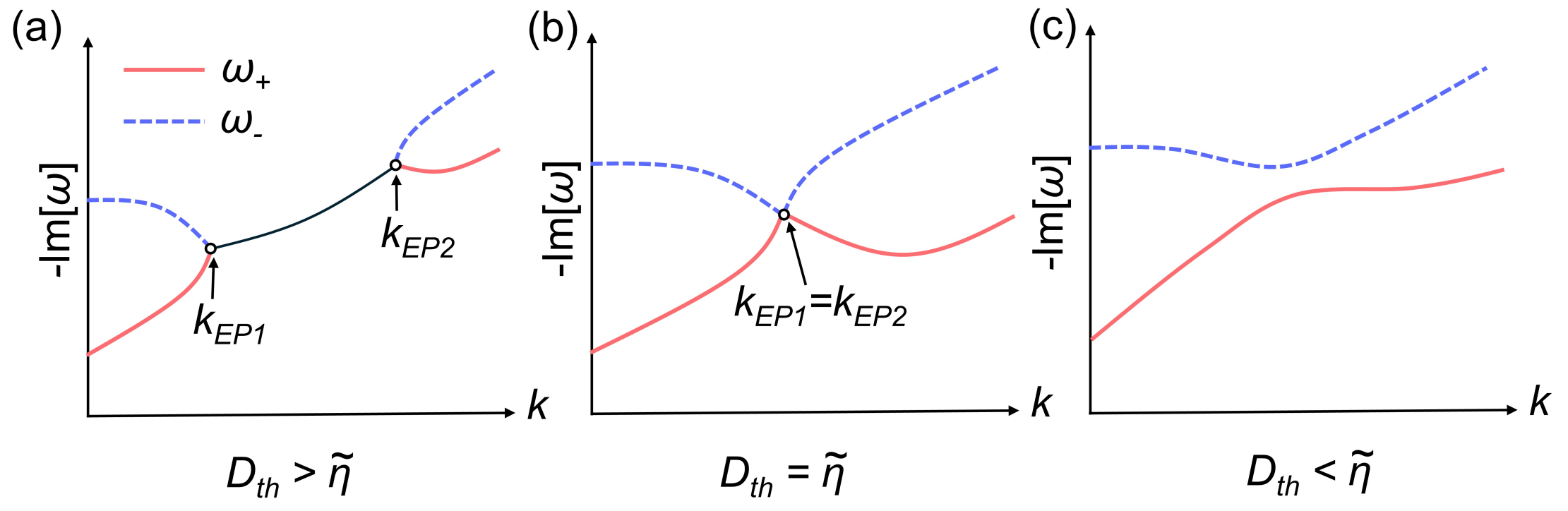

FIG. 3. Imaginary parts of the thermal-wave dispersions for (a) $D_{\mathrm{th}} > \tilde{\eta}$, where a finite hydrodynamic window exists; (b) $D_{\mathrm{th}} = \tilde{\eta}$, where the two exceptional points coalesce; and (c) $D_{\mathrm{th}} < \tilde{\eta}$, where hydrodynamic transport cannot be supported.

*Ultrafast second sound oscillations* – The exceptional-point framework developed above provides not only a criterion for identifying hydrodynamic transport, but also a predictive tool for calculating experimentally observable thermal dynamics. We demonstrate this capability by first-principles modeling of ultrafast second-sound oscillations in transient thermal grating (TTG) experiments with no empirical parameters. As shown in Fig. 4(a), TTG uses two interfering pump beams at the sample surface to generate a spatially periodic surface heat flux, with the grating period setting the thermal wavevector. The subsequent dynamics of the thermal fields are calculated from the effective evolution equation $|\psi(t)\rangle = U(t)|\psi(0)\rangle$, in which the initial state $|\psi(0)\rangle$ evolves under the non-Hermitian time-evolution operator $U(t) = \exp(-i\mathcal{H}t)$. Evaluating this operator yields:

$$U(t) = e^{-\Gamma_\theta t} \begin{bmatrix} \cos(\Omega_\theta t) - \frac{\zeta_\theta}{\Omega_\theta}\sin(\Omega_\theta t) & -i\frac{\tilde{v}_0 k}{\Omega_\theta}\sin(\Omega_\theta t) \\ -i\frac{\tilde{v}_0 k}{\Omega_\theta}\sin(\Omega_\theta t) & \cos(\Omega_\theta t) + \frac{\zeta_\theta}{\Omega_\theta}\sin(\Omega_\theta t) \end{bmatrix} \quad (18)$$

where $\zeta_\theta = \sqrt{(\tilde{v}_0 k)^2 - \Omega_\theta^2}$. For a thermalizing heat source, only energy is injected into the system with no net momentum flux (*i.e.,* no net forces). Therefore, at $t = 0$, we can set $|\psi(0)\rangle = [1,0]^T$, and the temperature oscillation is therefore:

$$\theta(t) = U_{11}(t) = e^{-\Gamma_\theta t}\left[\cos(\Omega_\theta t) - \frac{\zeta_\theta}{\Omega_\theta}\sin(\Omega_\theta t)\right] \quad (19)$$

Fig. 4(b) shows a typical second sound oscillation in graphite, where the grating size is $L = 10$ μm at 100 K. The transient lattice cooling at ~1.4 ns is well captured, consistent with TTG measurements [29]. The transient cooling is due to the constructive interference of collective thermal waves excited by the grating heat source, such that a $\pi$-phase inversion occurs relative to the initial temperature. The excellent agreement in both the oscillation timescale and the cooling minimum demonstrates that the non-Hermitian framework captures the essential collective transport dynamics governing the hydrodynamic second sound. Another advantage of modeling second sound using our non-Hermitian framework is that this method avoids the heavy computational demand of inverting the full scattering matrix [35]. Compared with the TTG experiments[29] and the full scattering simulations[35], the bump at 2~3 ns is weaker because our theory has neglected high-order terms of $\omega\tau_\mu$ (*i.e.* high-order time derivatives), and detailed selection rules for phonon-phonon interactions are not considered in our theory [37]. In Fig. 4(c), we calculated the temperature-dependent TTG responses using the non-Hermitian theory. At the grating period of 10 μm, second sound only exists from 75 K to 125 K, agreeing well with TTG experiments [29]. At the fixed grating period of 10 μm, the transient cooling is absent below 50 K or above 150 K. Because in the quasi-ballistic regime, phonon momentum coupling is too weak to support collective wavelike transport, and the dominating Umklapp scattering at elevated temperatures dissipates the phonon momentum strongly. In either case, the APT symmetry is preserved where the second sound is not allowed.

To further connect the non-Hermitian phase-transition picture with ultrafast TTG experiments, Fig. 4(d) maps the strength of the second sound using the negative temperature dip as the indicator. The EP pair derived from the previous sections sets the upper and lower bounds for the grating sizes, between which second sound can be observed. The lower and the upper bounds of the hydrodynamic window are defined as $L_{H,min} = 2\pi / k_{EP2}$ and $L_{H,max} = 2\pi/k_{EP1}$. Remarkably, the region exhibiting pronounced second-sound signatures is well confined between $L_{H,min}$ and $L_{H,max}$, demonstrating that the EP pair quantitatively captures the onset conditions of phonon hydrodynamics. At 50 K, $L_{H,min}$ is predicted to be approximately 12 μm, exceeding the 10 μm grating period used in Ref. [29] and thereby explaining the absence of second sound under this condition. The predicted hydrodynamic window broadens as the temperature decreases, consistent with the viscous-heat-equation prediction of Simoncelli *et al* [43]. This broadening arises because weaker Umklapp scattering reduces the momentum-dissipation rate $\tilde{\gamma}$, producing a rapid increase in $L_{\mathrm{H,max}}$. In addition to $\tilde{\gamma}$, $L_{\mathrm{H,min}}$ depends on the difference between the thermal and momentum diffusivities and therefore increases more slowly than $L_{\mathrm{H,max}}$ with decreasing temperature.

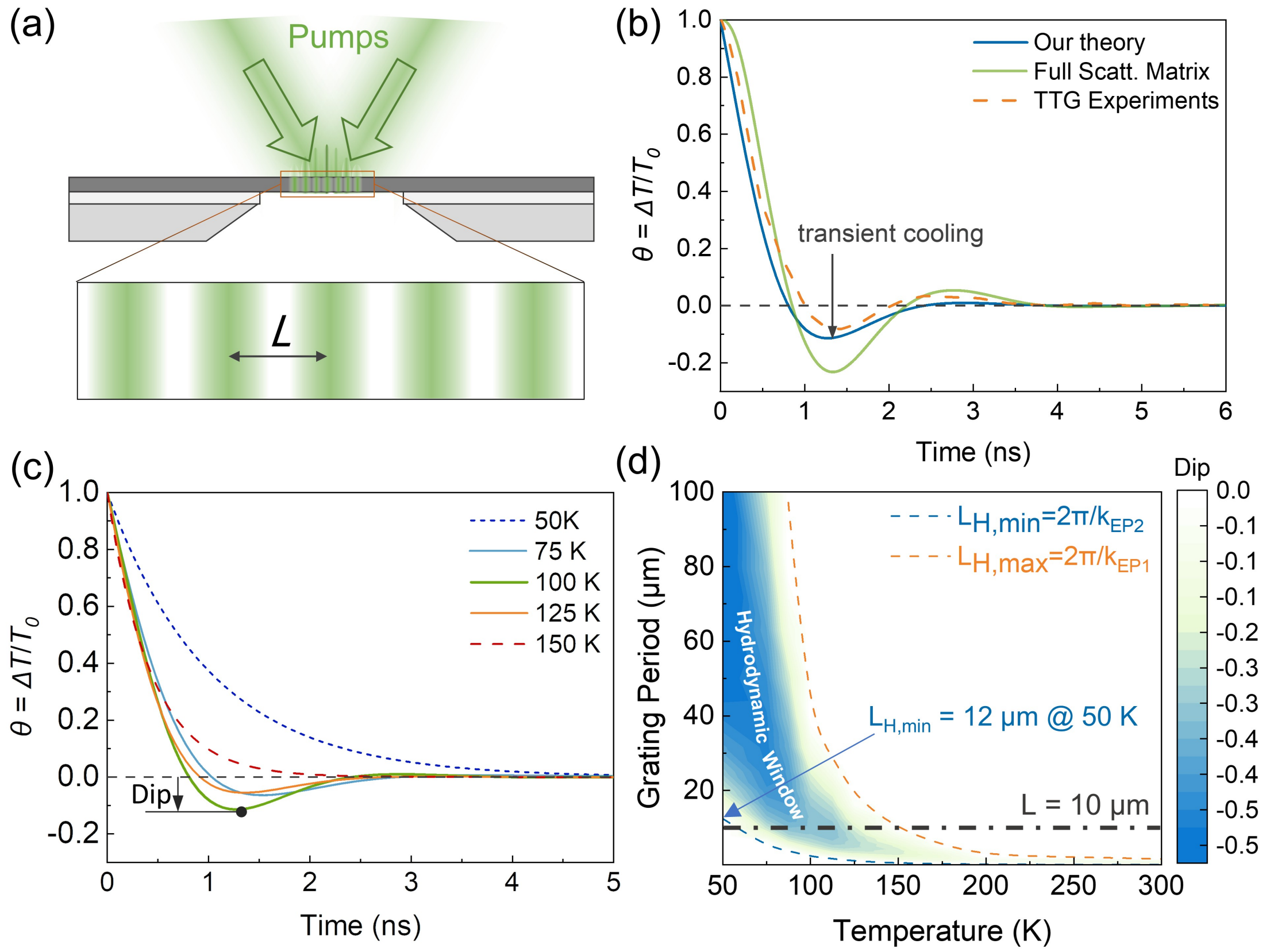

FIG. 4. (a) Schematic of TTG experiments. (b) Wavelike oscillations in $\theta(t)$ using Eq. (18) (Our theory), compared with full scattering matrix calculations, and TTG experiments[29] at 100 K and $L$ =10 μm. (c) Temperature-dependent $\theta(t)$ with $L$ =10 μm. (d) Contour of the first dip and the temperature-dependent hydrodynamic window.

*Conclusion* – We have demonstrated that the emergence of hydrodynamic second sound can be interpreted as a non-Hermitian phase transition inherent to collective phonon transport. Starting from the PBTE, we derived an effective Schrödinger equation for thermal waves, wherein the governing Hamiltonian is non-Hermitian and obeys anti-parity-time (APT) symmetry. Within this theoretical framework, the onset of second sound originates from spontaneous breaking of APT symmetry, and a pair of exceptional points (EPs) quantitatively define the lower and upper bounds of the hydrodynamic window. Our theory further establishes a material-specific criterion for the existence of hydrodynamic transport: a finite hydrodynamic window can be sustained only when the effective thermal diffusivity, contributed by both the kinetic diffusion and collective motion, exceeds the phononic momentum diffusivity. These findings establish a symmetry-based framework for understanding collective energy transport across diffusive, hydrodynamic, and quasiballistic regimes, while providing quantitative design principles for identifying the material, temperature, and length scales that enable wavelike heat transport in low-dimensional materials and functional devices. In the future, it will be valuable to extend this non-Hermitian framework to electron hydrodynamics [47] and coupled electron-phonon hydrodynamic transport [48].

## Methods

Further details can be found in supplementary materials, regarding the derivation of the linear equation for temperature and collective drift fields, effective Schrödinger equation of thermal waves, symmetrized non-Hermitian Hamiltonian, relation to the viscous heat equation, dissipation mechanisms, and ab initio phonon calculation details.

*Acknowledgments*

*Acknowledgments*. X.Q. acknowledges funding support from the National Key R & D Program of China (No. 2022YFA1203100). R.Y., N.W, J.C., and X.X. acknowledge the support from the National Natural Science Foundation of China (NSFC, Grant Nos. T2588301, 12425414, 12304571, and 12174286). X.Q., J.C. and R.Y. conceived the ideas, X.Q., G.Q., and N.W developed the results presented in this work, X.Q. and N.W. drafted the manuscript, and all authors edited the paper. X.Q., J.C. and R.Y. supervised this project.

# SUPPLEMENTAL INFORMATION:

## Emergence of Phonon Hydrodynamics as a Non-Hermitian Symmetry-Braking Phase Transition

Xin Qian[1], Guanda Quan[1], Ning Wang [2,*], Yaoming Chu[2], Run Hu[1], Te-Huan Liu[1], Xiangfan Xu[3], Jianming Cai[2,†], and Ronggui Yang[4,‡]

[1]School of Energy and Power Engineering, Huazhong University of Science and Technology, Wuhan 430074, China

[2] School of Physics, International Joint Laboratory on Quantum Sensing and Quantum Metrology, Hubei Key Laboratory of Gravitation and Quantum Physics, Institute for Quantum Science and Engineering, Center for Intelligence and Quantum Science (CIQS), Huazhong University of Science and Technology, Wuhan 430074, China

[3]Center for Phononics and Thermal Energy Science, School of Physics Science and Engineering, Tongji University, Shanghai 200092, China

[4]College of Mechanics and Engineering Science, Peking University, Beijing, 100871, China

*Contact author: ningwang@hust.edu.cn

†Contact author: jianmingcai@hust.edu.cn

‡Contact author: ronggui@pku.edu.cn

**List of Contents:**

## Note 1: Linear equation of $\boldsymbol{\Delta T}$ and $\boldsymbol{u}$

To simultaneously solve the temperature rise and the drift velocity, energy and momentum conservation constraints are considered to obtain closed-form equations. By multiplying $\hbar\omega_\mu$ on both sides of Eq. (1), and summing over all modes $\mu$ over the Brillouin zone, the continuity equation of heat energy can be derived:

$$\frac{\partial e}{\partial t}+\nabla\cdot\boldsymbol{J}_q=\sum_\mu\frac{C_\mu\Delta T-\hbar\omega_\mu n_\mu}{\tau_\mu} \tag{S1}$$

where $e=\sum_\mu\hbar\omega_\mu n_\mu$ is the nonequilibrium energy density, $\boldsymbol{J}_q=\sum_\mu\hbar\omega_\mu\boldsymbol{v}_\mu n_\mu$ is the heat flux. The right-hand side of Eq. (S1) must be zero to ensure energy conservation:

$$\sum_\mu\frac{C_\mu\Delta T-\hbar\omega_\mu n_\mu}{\tau_\mu}=0 \tag{S2}$$

Similarly, the continuity equation for phonon momentum can be obtained by multiplying $\hbar\boldsymbol{q}_\mu$ and summing over all modes:

$$\frac{\partial\boldsymbol{P}}{\partial t}+\nabla\cdot\boldsymbol{J}_{\boldsymbol{P}}=\sum_\mu\hbar\boldsymbol{q}_\mu\frac{N_\mu-N_\mu^0(T)}{\tau_\mu^R}+\sum_\mu\hbar\boldsymbol{q}_\mu\frac{N_\mu-N_\mu^D(T)}{\tau_\mu^N} \tag{S3}$$

where $\boldsymbol{P}=\sum_\mu\hbar\boldsymbol{q}_\mu n_\mu$ and $\boldsymbol{J}_{\boldsymbol{P}}=\sum_\mu\hbar\boldsymbol{q}_\mu\boldsymbol{v}_\mu n_\mu$ is the momentum flux. Since N-scattering events conserve momentum, we have:

$$\sum_\mu\hbar\boldsymbol{q}_\mu\frac{N_\mu-N_\mu^D(T,\boldsymbol{u})}{\tau_\mu^N}=\sum_\mu\hbar\boldsymbol{q}_\mu\frac{n_\mu-\left(\frac{C_\mu\Delta T}{\hbar\omega_\mu}+\frac{C_\mu T_0}{\hbar\omega_\mu}\cdot\frac{\boldsymbol{q}_\mu\cdot\boldsymbol{u}}{\omega_\mu}\right)}{\tau_\mu^N}=0 \tag{S4}$$

Substituting Eq. (2) into Eq. (S2) and Eq. (S4), we can obtain the linear system Eq. (3) in the main text.

## Note 2: Emergence of SE under the slow-varying approximation

We now show that while the microscopic transport picture of phonons remains in the quasiparticle regime, the macroscopic fields ($\Delta T$ and $u$) can be described by an effective Schrödinger equation (SE). The slowing-varying approximation is assumed: $\omega\tau_\mu\ll 1$. This is valid because thermal responses in TTG are in a few nanoseconds (GHz range), while the phonon lifetimes are on the order of picoseconds. However, since the grating size is in a few micrometers, on the same order of mean free displacements, the long-wavelength approximation is not invoked in our derivation to properly consider quasi-ballistic effects. As we shall see in the derivations, the quasi-ballistic effect can be effectively incorporated by correcting phonon properties using mode-resolved suppression functions.

To derive the effective Schrödinger equation, the elements of the linear matrix are evaluated. The matrix element $\mathbb{A}_{TT}$ can be calculated as:

$$\begin{aligned}\mathbb{A}_{TT}=\sum_\mu\frac{C_\mu}{\tau_\mu}(1-\chi_\mu)&=\frac{1}{2}\left[\sum_\mu\frac{C_\mu}{\tau_\mu}(1-\chi_\mu)+\sum_\mu\frac{C_{-\mu}}{\tau_{-\mu}}(1-\chi_{-\mu})\right]\\&=\sum_\mu\frac{C_\mu}{\tau_\mu}\left(1-\frac{\chi_\mu+\chi_{-\mu}}{2}\right)\end{aligned} \tag{S5}$$

Note that we have symmetrized the expression by summing over phonon modes with opposite wavevectors $\mu = (\boldsymbol{q}, s)$ and $-\mu = (-\boldsymbol{q}, s)$. Due to the inversion symmetry, $C_{-\mu} = C_\mu, \tau_{-\mu} = \tau_\mu$ are even properties. Substituting $\chi_\mu = \left(1 + i\omega\tau_\mu - i\boldsymbol{k}\cdot\boldsymbol{F}_\mu\right)^{-1}$ into Eq. (S5), we can obtain:

$$\begin{aligned}\mathbb{A}_{TT} &= \sum_\mu \frac{C_\mu}{\tau_\mu}\left[1 - \frac{1 + i\omega\tau_\mu}{\left(1 + i\omega\tau_\mu\right)^2 + \left(\boldsymbol{k}\cdot\boldsymbol{F}_\mu\right)^2}\right] \approx \sum_\mu \frac{C_\mu}{\tau_\mu}\frac{i\omega\tau_\mu + \left(\boldsymbol{k}\cdot\boldsymbol{F}_\mu\right)^2}{1 + \left(\boldsymbol{k}\cdot\boldsymbol{F}_\mu\right)^2} \\ &= \sum_\mu \frac{C_\mu}{1 + \left(\boldsymbol{k}\cdot\boldsymbol{F}_\mu\right)^2} i\omega + \sum_\mu \frac{C_\mu(\boldsymbol{k}\cdot\boldsymbol{v}_\mu)(\boldsymbol{k}\cdot\boldsymbol{F}_\mu)}{1 + \left(\boldsymbol{k}\cdot\boldsymbol{F}_\mu\right)^2}\end{aligned} \tag{S6}$$

We have only invoked the approximation $\omega\tau_\mu \ll 1$ in simplifying Eq. (S6) since we only keep the leading term $\mathcal{O}(\omega\tau_\mu)$. In contrast, all orders of $\mathcal{O}\left(\left|\boldsymbol{k}\cdot\boldsymbol{F}_\mu\right|^n\right)$ have been included in the derivation; therefore, this derivation is still valid in small spatial scales with quasiballistic effects. The matrix element is therefore rewritten as:

$$\mathbb{A}_{TT} \approx i\omega \sum_\mu \frac{C_\mu}{1 + \left(\boldsymbol{k}\cdot\boldsymbol{F}_\mu\right)^2} + k^\alpha \sum_\mu \frac{C_\mu v_\mu^\alpha F_\mu^\beta}{1 + \left(\boldsymbol{k}\cdot\boldsymbol{F}_\mu\right)^2} k^\beta = i\omega\tilde{C} + k^\alpha \tilde{\kappa}^{\alpha\beta} k^\beta \tag{S7}$$

where the superscripts $\alpha, \beta$ represent the indices of Cartesian coordinates, and we have used the Einstein summation notation for these directional indices. Here, we can define the suppression function as

$$S_\mu(\boldsymbol{k}) = \frac{1}{1 + \mathrm{Kn}_\mu^2} = \frac{1}{1 + \left(\boldsymbol{k}\cdot\boldsymbol{F}_\mu\right)^2} \tag{S8}$$

where $\mathrm{Kn}_\mu = \left|\boldsymbol{k}\cdot\boldsymbol{F}_\mu\right|$ is the Knudsen number. This suppression function is identical to the results derived by Minnich [28], allowing for mode-resolved correction to thermal properties in anisotropic materials. The effective specific heat and the effective thermal conductivity are defined as $\tilde{C} = \sum_\mu C_\mu S_\mu$, $\tilde{\kappa}^{\alpha\beta} = \sum_\mu C_\mu v_\mu^\alpha F_\mu^\beta S_\mu$. Note that the effective specific heat is also corrected by the suppression function $S_\mu$. However, this does not mean that the equilibrium specific heat is size-dependent; rather, $\tilde{C}$ here should be understood as an effective value capturing the transient energy storage kinetics due to the phonon nonequilibrium populations and nonlocal transport effects (*i.e.* wavevector dependence) at micro/nanoscales. Such corrections are only significant at a large $\mathrm{Kn}_\mu$. Similarly, we can calculate the coupling sub-matrix block $\mathbb{A}_{cp}$ in Eq. (3):

$$\begin{aligned}\mathbb{A}_{cp} &= -\sum_\mu \frac{C_\mu \boldsymbol{q}_\mu}{\omega_\mu \tau_\mu^N}\left(\chi_\mu - \chi_{-\mu}\right)/2 = \left(\sum_\mu \eta_\mu \frac{C_\mu \boldsymbol{q}_\mu \boldsymbol{v}_\mu}{\omega_\mu \tau_\mu^N}\frac{1}{1 + \left(\boldsymbol{k}\cdot\boldsymbol{F}_\mu\right)^2}\right)\cdot(-i\boldsymbol{k}) \\ &= \widetilde{\boldsymbol{\Pi}}\cdot(-i\boldsymbol{k})\end{aligned} \tag{S9}$$

The matrix elements of the drifting momentum transport coefficient $\widetilde{\boldsymbol{\Pi}}$ is expressed as $\widetilde{\Pi}^{\alpha\beta} = \sum_\mu \eta_\mu \frac{C_\mu q_\mu^\alpha v_\mu^\beta}{\omega_\mu} S_\mu$.

Finally, the sub-matrix block $\mathbb{A}_{uu}$ is calculated as:

$$\begin{aligned}\mathbb{A}_{uu}^{\alpha\beta} &= T_0 \sum_\mu \frac{C_\mu q_\mu^\alpha q_\mu^\beta}{\omega_\mu^2 \tau_\mu^N}\left(1 - \eta_\mu \frac{\chi_\mu + \chi_{-\mu}}{2}\right) \\ &\approx T_0 \sum_\mu \frac{C_\mu q_\mu^\alpha q_\mu^\beta}{\omega_\mu^2 \tau_\mu^N} S_\mu \left[\rho_\mu + \left(2 - \eta_\mu\right) i\omega\tau_\mu + k^\xi F_\mu^\xi k^\zeta F_\mu^\zeta\right]\end{aligned} \tag{S10}$$

where $\rho_\mu = 1 - \eta_\mu = \tau_\mu/\tau_\mu^R$ is the relative strength of the R-scattering event. The matrix element $A_{uu}^{\alpha\beta}$ can be further reorganized as follows:

$$\mathbb{A}_{uu}^{\alpha\beta} = T_0\left(\tilde{\Gamma}^{\alpha\beta} + \tilde{A}^{\alpha\beta} i\omega + \tilde{\mathrm{M}}^{\alpha\beta\xi\zeta} k^\xi k^\zeta\right) \quad \text{(S11)}$$

where $\tilde{\Gamma}^{\alpha\beta} = \sum_\mu \frac{C_\mu q_\mu^\alpha q_\mu^\beta}{\omega_\mu^2} \frac{\tau_\mu}{\tau_\mu^N \tau_\mu^R} S_\mu$ , $\tilde{A}^{\alpha\beta} = \sum_\mu \frac{C_\mu q_\mu^\alpha q_\mu^\beta}{\omega_\mu^2} \eta_\mu (2 - \eta_\mu) S_\mu$ and $\tilde{\mathrm{M}}^{\alpha\beta\xi\zeta} = \sum_\mu \eta_\mu \frac{C_\mu q_\mu^\alpha q_\mu^\beta}{\omega_\mu^2} v_\mu^\xi F_\mu^\zeta S_\mu$ are the momentum dissipation rate, specific momentum intensity, and the phonon viscosity tensor. Substituting Eqs. (S7), (S9), (S10) into Eq. (3), and replacing $\Delta T$ to normalized temperature $\theta = \Delta T / T_0$, we can obtain:

$$i\omega \begin{bmatrix} \theta \\ u^\beta \end{bmatrix} = \begin{bmatrix} -k^\alpha \frac{\tilde{\kappa}^{\alpha\beta}}{\tilde{C}} k^\beta & -ik^\alpha \frac{\tilde{\Pi}^{\alpha\beta}}{\tilde{C}} \\ -ik^\alpha \left(\tilde{A}^{-1}\tilde{\Pi}\right)^{\alpha\beta} & -\left[\left(\tilde{A}^{-1}\tilde{\mathrm{M}}\right)^{\alpha\beta\xi\zeta} k^\xi k^\zeta + \left(\tilde{A}^{-1}\tilde{\Gamma}\right)^{\alpha\beta}\right] \end{bmatrix} \begin{bmatrix} \theta \\ u^\beta \end{bmatrix} \quad \text{(S12)}$$

where $\left(\tilde{A}^{-1}\tilde{\Pi}\right)^{\alpha\beta} = \left(\tilde{A}^{-1}\right)^{\alpha\gamma} \tilde{\Pi}^{\gamma\beta}$ , $\left(\tilde{A}^{-1}\tilde{\Gamma}\right)^{\alpha\beta} = \left(\tilde{A}^{-1}\right)^{\alpha\gamma} \tilde{\Gamma}^{\gamma\beta}$ , and $\left(\tilde{A}^{-1}\tilde{\mathrm{M}}\right)^{\alpha\beta\xi\zeta} = \left(\tilde{A}^{-1}\right)^{\alpha\gamma} \tilde{\mathrm{M}}^{\gamma\beta\xi\zeta}$ are the matrix multiplications. By performing inverse Fourier transform ($i\omega \to \partial/\partial t$) and multiplying an imaginary unit on both sides, we obtained the unsymmetrized effective Schrödinger equation:

$$i\frac{\partial}{\partial t}|\phi\rangle = \mathcal{H}_A(\boldsymbol{k})|\phi\rangle \quad \text{(S13)}$$

where $|\phi\rangle = \left[\theta, u^\beta\right]^T$ is the unsymmetrized state vector, and $\mathcal{H}_A(\boldsymbol{k})$ is the unsymmetrized Hamiltonian of thermal waves:

$$\mathcal{H}_A(\boldsymbol{k}) = \begin{bmatrix} -k^\alpha \frac{\tilde{\kappa}^{\alpha\beta}}{\tilde{C}} k^\beta & k^\alpha \frac{\tilde{\Pi}^{\alpha\beta}}{\tilde{C}} \\ k^\alpha \left(\tilde{A}^{-1}\tilde{\Pi}\right)^{\alpha\beta} & -\left[\left(\tilde{A}^{-1}\tilde{\mathrm{M}}\right)^{\alpha\beta\xi\zeta} k^\xi k^\zeta + \left(\tilde{A}^{-1}\tilde{\Gamma}\right)^{\alpha\beta}\right] \end{bmatrix} \quad \text{(S14)}$$

## Note 3: Symmetrized Hamiltonian

To clearly probe the symmetry of the thermal wave system, $\mathcal{H}_A$ needs to be symmetrized. For mathematical simplicity, we first symmetrize the Hamiltonian in the 1D case. The 1D effective Schrödinger equation is:

$$i\frac{\partial}{\partial t}\begin{bmatrix} \theta \\ u \end{bmatrix} = \begin{bmatrix} -i\frac{\tilde{\kappa}}{\tilde{C}} k^2 & \frac{\tilde{\Pi}}{\tilde{C}} k \\ \frac{\tilde{\Pi}}{\tilde{A}} k & -i\left(\frac{\tilde{\mathrm{M}}}{\tilde{A}} k^2 + \frac{\tilde{\Gamma}}{\tilde{A}}\right) \end{bmatrix} \begin{bmatrix} \theta \\ u \end{bmatrix} \quad \text{(S15)}$$

The effective Hamiltonian can be symmetrized using the transformation:

$$|\psi\rangle = \mathcal{S}|\phi\rangle = \begin{bmatrix} \sqrt{\tilde{C}} & 0 \\ 0 & \sqrt{\tilde{A}} \end{bmatrix} \begin{bmatrix} \theta \\ u \end{bmatrix} = \begin{bmatrix} \sqrt{\tilde{C}}\theta \\ \sqrt{\tilde{A}}u \end{bmatrix} \quad \text{(S16)}$$

Under such scaling of temperature rise and drift velocity, the effective Schrödinger equation is symmetrized as:

$$i\frac{\partial}{\partial t}|\psi\rangle = \mathcal{H}(k)|\psi\rangle \quad \text{(S17)}$$

Here $\mathcal{H}(k)$ is the symmetrized Hamiltonian defined in Eq. (3) of the main text. The coefficients in the Hamiltonian $\tilde{\alpha}$, $\tilde{\eta}$, and $\tilde{\gamma}$ are correspondingly the effective thermal diffusivity, effective momentum diffusivity, and specific momentum dissipation rate; $\tilde{v}_0$ is the inviscid second sound speed. These coefficients are calculated as:

$$\tilde{\alpha} = \tilde{\kappa}/\tilde{C} = \frac{\sum_\mu C_\mu v_\mu^x F_\mu^x S_\mu}{\sum_\mu C_\mu S_\mu} \tag{S18}$$

$$\tilde{\gamma} = \frac{\sum_\mu \frac{C_\mu (q_\mu^x)^2}{\omega_\mu^2} \frac{\tau_\mu}{\tau_\mu^N \tau_\mu^R} S_\mu}{\sum_\mu \frac{C_\mu (q_\mu^x)^2}{\omega_\mu^2} \eta_\mu (2-\eta_\mu) S_\mu} \tag{S19}$$

$$\tilde{\eta} = \tilde{\mathrm{M}}/\tilde{A} = \frac{\sum_\mu \eta_\mu^2 \frac{C_\mu}{\omega_\mu^2} (q_\mu^x)^2 v_\mu^x F_\mu^x S_\mu}{\sum_\mu \frac{C_\mu (q_\mu^x)^2}{\omega_\mu^2} \eta_\mu (2-\eta_\mu) S_\mu} \tag{S20}$$

$$\tilde{v}_0 = \frac{\tilde{\Pi}}{\sqrt{\tilde{A}\tilde{C}}} = \frac{\sum_\mu \eta_\mu \frac{C_\mu q_\mu^x v_\mu^x}{\omega_\mu} S_\mu}{\sqrt{\left(\sum_\mu \frac{C_\mu (q_\mu^x)^2}{\omega_\mu^2} \eta_\mu (2-\eta_\mu) S_\mu\right)\left(\sum_\mu C_\mu S_\mu\right)}} \tag{S21}$$

At the bulk limit $\mathrm{Kn}_\mu \to 0$ and the N-scattering limit $\eta_\mu \to 1$, $\tilde{v}_0$ is reduced to:

$$\lim_{\substack{\eta_\mu \to 1 \\ \mathrm{Kn}_\mu \to 0}} \tilde{v}_0 = v_0 = \frac{\sum_\mu \frac{C_\mu q_\mu^x v_\mu^x}{\omega_\mu}}{\sqrt{C\left(\sum_\mu \frac{C_\mu (q_\mu^x)^2}{\omega_\mu^2}\right)}} \tag{S22}$$

Eq. (S22) is identical to the derivation by Hardy and Ding [29,30] assuming the limit of strong N-scattering. In the multi-dimensional case, the state vector is $|\psi\rangle = \left[\sqrt{\tilde{C}}\theta, \tilde{\boldsymbol{A}}^{1/2} \cdot \boldsymbol{u}\right]^T$, and the symmetrized Hamiltonian is:

$$\mathcal{H}(\boldsymbol{k}) = \begin{bmatrix} \boldsymbol{H}_{\theta\theta} & \boldsymbol{H}_{\theta u} \\ \boldsymbol{H}_{\theta u}^T & \boldsymbol{H}_{uu} \end{bmatrix} \tag{S23}$$

The sub-matrices expressed in coordinate indices are written as:

$$H_{\theta\theta}^{\alpha\beta} = -ik^\alpha \frac{\tilde{\kappa}^{\alpha\beta}}{\tilde{C}} k^\beta \tag{S24}$$

$$H_{\theta u}^{\alpha\beta} = k^\alpha \tilde{v}_0^{\beta\alpha} \tag{S25}$$

$$H_{uu}^{\alpha\beta} = -i\left[\left(\tilde{A}^{-1}\tilde{\mathrm{M}}\right)^{\alpha\beta\xi\zeta} k^\xi k^\zeta + \left(\tilde{A}^{-1}\tilde{\Gamma}\right)^{\alpha\beta}\right] \tag{S26}$$

$\tilde{v}_0^{\alpha\beta}$ is the inviscid phase velocity tensor:

$$\tilde{v}_0^{\alpha\beta} = \frac{1}{\sqrt{\tilde{C}}} \left(\tilde{A}^{-1/2}\right)^{\alpha\xi} \tilde{\Pi}^{\xi\beta} \tag{S27}$$

Due to the inversion symmetry, it can be shown that $\tilde{\boldsymbol{A}}$ is a diagonal matrix, $\tilde{A}^{\alpha\beta} = \delta_{\alpha\beta}\sum_{\mu}\frac{C_{\mu}(q_{\mu}^{\alpha})^2}{\omega_{\mu}^2}\eta_{\mu}(2-\eta_{\mu})S_{\mu}$, with $\delta_{\alpha\beta}$ denoting the Kronecker delta. The tensor $\tilde{\boldsymbol{A}}^{-1/2}$ is therefore defined as $\left(\tilde{A}^{-1/2}\right)^{\alpha\xi} = \delta_{\alpha\xi}\Big/\sqrt{\sum_{\mu}\frac{C_{\mu}(q_{\mu}^{\alpha})^2}{\omega_{\mu}^2}\eta_{\mu}(2-\eta_{\mu})S_{\mu}}$.

We can now check the symmetry of the effective Hamiltonian. Under time reversal operation $\mathcal{T}$, not only is the direction of time flipped $t \to -t$, but also the collective phonon drift velocity is reversed: $u \to -u$. Hence, the time-reversed Hamiltonian is calculated as:

$$\mathcal{T}\mathcal{H}(k)\mathcal{T}^{-1} \coloneqq \sigma_z\mathcal{H}^*(k)\sigma_z = -\mathcal{H}(k) \tag{S28}$$

where $\sigma_z$ is the Pauli matrix in the $z$-direction. Under parity operation $\mathcal{P}: x \to -x, k \to -k, u \to -u$, the Hamiltonian is symmetric:

$$\mathcal{P}\mathcal{H}(k)\mathcal{P}^{-1} \coloneqq \sigma_z\mathcal{H}(-k)\sigma_z = \mathcal{H}(k) \tag{S29}$$

The dynamical system is anti-symmetric under the combined $\mathcal{PT}$ transformation: $(\mathcal{PT})\mathcal{H}(\mathcal{PT})^{-1} = -\mathcal{H}$. We therefore proved the APT symmetry of the Hamiltonian, *i.e.,* the $\mathcal{PH}$ anti-commutates with $\mathcal{H}$:

$$\{\mathcal{PT}, \mathcal{H}\} = 0 \tag{S30}$$

## Note 4: Viscous heat equation in the local transport limit

In this appendix, we show that the effective SE can be simplified to the VHE in the local transport limit. In each transport coefficient ($\tilde{C}, \tilde{\kappa}^{\alpha\beta}, \tilde{\Pi}^{\alpha\beta}, \tilde{\Gamma}^{\alpha\beta}, \tilde{M}^{\alpha\beta\xi\zeta}$) in Table I, there is a suppression function $S_{\mu}(\boldsymbol{k}) = 1/\left[1 + \left(\boldsymbol{k}\cdot\boldsymbol{F}_{\mu}\right)^2\right]$ depending on thermal wavevectors. Therefore, nonlocal effects are embedded in the effective SE of thermal waves. In weakly ballistic cases, the nonlocal effects can be neglected, which corresponds to a "mean field" treatment of the suppression function where the term $\boldsymbol{k}\cdot\boldsymbol{F}_{\mu}$ is lumped into a mean Knudsen number as $\overline{\mathrm{Kn}}_{\mu} = |\boldsymbol{F}_{\mu}|/L_c$, with $L_c$ being the characteristic length. For example, in hydrodynamic transport in nanoribbons, $L_c$ can be taken as the width of the ribbon. After such a mean field treatment, $S_{\mu}(\boldsymbol{k})$ simply becomes a correction factor $\bar{S}_{\mu} = 1/\left(1 + \overline{\mathrm{Kn}}_{\mu}^2\right)$ independent of $\boldsymbol{k}$. This kind of mean-field treatment is only valid in the weak-ballistic regime with $\mathrm{Kn}_{\mu} < 1$, where the nonlocal transport effects are neglected. Applying inverse Fourier transform ($i\omega \to \frac{\partial}{\partial t}, -ik^{\alpha} \to \frac{\partial}{\partial x_{\alpha}}$), Eq. (S12) can be transformed to the VHE:

$$\tilde{C}\frac{\partial\theta}{\partial t} + \tilde{\Pi}^{\beta\alpha}\frac{\partial u^{\beta}}{\partial x_{\alpha}} - \tilde{\kappa}^{\alpha\beta}\frac{\partial^2\theta}{\partial x_{\alpha}\partial x_{\beta}} = 0 \tag{S31}$$

$$\tilde{A}^{\alpha\beta}\frac{\partial u^{\beta}}{\partial t} + \tilde{\Pi}^{\alpha\beta}\frac{\partial\theta}{\partial x_{\beta}} - \tilde{M}^{\alpha\beta\xi\zeta}\frac{\partial^2 u^{\beta}}{\partial x_{\xi}\partial x_{\zeta}} = -\tilde{\Gamma}^{\alpha\beta}u^{\beta} \tag{S32}$$

The above equations are formally identical to the original derivation of VHE starting from the relaxon theory [31]. Comparing Eq. (S31) with the energy conservation equation $C\frac{\partial T}{\partial t} + \frac{\partial}{\partial x_{\alpha}}J_q^{\alpha} = 0$, we can see that the energy flux $J_q^{\alpha}$ can be decomposed into two parts:

$$J_q^\alpha = J_{qH}^\alpha + J_{qK}^\alpha \quad \text{(S33)}$$

Here $J_{qH}^\alpha$ and $J_{qK}^\alpha$ denote the heat flux due to collective drifting and kinetic random motion of phonons, respectively:

$$J_{qH}^\alpha = T_0 \tilde{\Pi}^{\beta\alpha} u^\beta \quad \text{(S34)}$$

$$J_{qK}^\alpha = -\tilde{\kappa}^{\alpha\beta} \frac{\partial T}{\partial x_\beta} \quad \text{(S35)}$$

In deriving the effective SE, the only approximation we make is the slow-varying limit, with $\omega\tau_\mu \ll 1$, and keeps all orders of $\mathcal{O}\left(\left|\boldsymbol{k}\cdot\boldsymbol{F}_\mu\right|^n\right)$. Therefore, the effective SE is valid across a wide length scale from the qusiballistic to the bulk diffusive limit, as long as the timescale of thermal excitation fluctuations is much larger than the phonon lifetimes. In addition, the effective SE is written in the wavevector space, allowing convenient determination of the second sound dispersions.

To establish the governing partial differential equation, we need to perform a mean-field treatment on the suppression function $S_\mu(\boldsymbol{k}) \rightarrow \bar{S}_\mu$ to eliminate the wavevector dependence in the transport coefficients ($\tilde{C}, \tilde{\kappa}^{\alpha\beta}, \tilde{\Pi}^{\alpha\beta}, \tilde{\Gamma}^{\alpha\beta}, \tilde{\mathrm{M}}^{\alpha\beta\xi\zeta}$). This simplification holds in the weak-ballistic case where the temperature field does not change steeply compared to the phonon mean free paths, and the system can still be effectively described by a set of PDEs with finite order differentiation of macroscopic fields $\theta$ and $\boldsymbol{u}$. It is easy to show that, applying the Fourier transform on the spatial coordinates of the VHE, one can restore the effective SE. Therefore, VHE can be regarded as a special case of the effective SE of thermal waves. We have summarized the relation and differences between the effective SE and the VHE in Table S1. Note that all numerical results on second sound dispersions presented in the main text are directly solved from the effective SE.

**Table S1. Comparison between the effective SE and VHE.**

<table>
<tr><td>Equation name</td><td>Effective Schrödinger equation (SE)</td><td>Viscous heat equation (VHE)</td></tr>
<tr><td>Formula</td><td>$i\frac{\partial}{\partial t}|\psi\rangle = \mathcal{H}(\boldsymbol{k})|\psi\rangle$<br>$|\psi\rangle = \left[\sqrt{\tilde{C}}\,\theta, \sqrt{\tilde{A}}\,u\right]^T$</td><td>$\tilde{C}\frac{\partial\theta}{\partial t} + \tilde{\Pi}^{\beta\alpha}\frac{\partial u^\beta}{\partial x_\alpha} - \tilde{\kappa}^{\alpha\beta}\frac{\partial^2\theta}{\partial x_\alpha \partial x_\beta} = 0$<br>$\tilde{A}^{\alpha\beta}\frac{\partial u^\beta}{\partial t} + \tilde{\Pi}^{\alpha\beta}\frac{\partial\theta}{\partial x_\beta} - \tilde{\mathrm{M}}^{\alpha\beta\xi\zeta}\frac{\partial^2 u^\beta}{\partial x_\xi \partial x_\zeta} = -\tilde{\Gamma}^{\alpha\beta}u^\beta$</td></tr>
<tr><td>Variable space</td><td>Time domain & wavevector space $(t, \boldsymbol{k})$</td><td>Time domain & real space $(t, \boldsymbol{r})$</td></tr>
<tr><td>Applicable condition</td><td>Slow-varying:<br>$\omega\tau_\mu \ll 1$</td><td>Slow-varying and local transport:<br>$\omega\tau_\mu \ll 1, \mathrm{Kn}_\mu < 1$</td></tr>
<tr><td>Relation</td><td colspan="2">• Both are simplifications of the PBTE<br>• For the weak-ballistic case, the effective SE can be reduced to VHE by replacing $S_\mu(\boldsymbol{k})$ to a mean-field $\bar{S}_\mu$<br>$S_\mu(\boldsymbol{k}) = \frac{1}{1+(\boldsymbol{k}\cdot\boldsymbol{F}_\mu)^2} \rightarrow \bar{S}_\mu = \frac{1}{1+(|\boldsymbol{F}_\mu|/L_c)^2}$<br>• VHE can be transformed back to the effective SE using Fourier transform</td></tr>
<tr><td>Usage of the Equation</td><td>• Symmetry Analysis<br>• Second sound dispersions<br>• Predicting EPs</td><td>• Simulation of complex structures</td></tr>
</table>

## Note 5: Dissipation mechanisms outside the hydrodynamic window

This section derives the scattering mechanisms in the short wavelength $k \gg k_{\mathrm{EP2}}$ and long wavelength $k \ll k_{EP1}$ regions outside the hydrodynamic window.

When $k \ll k_{EP1}$, $\Omega_\theta = i\sqrt{[\tilde{\gamma} - (\tilde{\alpha} - \tilde{\eta})k^2]^2/4 - (\tilde{v}_0 k)^2}$ is purely imaginary. Neglecting high-order terms $O(k^4)$ in the square root, we can approximately have:

$$\Omega_\theta \approx i\sqrt{\frac{\tilde{\gamma}^2}{4} - \left[\tilde{v}_0^2 + \frac{\tilde{\gamma}}{2}(\tilde{\alpha} - \tilde{\eta})\right]k^2} \approx i\frac{\tilde{\gamma}}{2} - i\left(\frac{\tilde{v}_0^2}{\tilde{\gamma}} + \frac{\tilde{\alpha} - \tilde{\eta}}{2}\right)k^2 \tag{S36}$$

In the second approximate equality, we have invoked the approximation $\sqrt{a^2 - bk^2} \approx a - \frac{b}{2a}k^2$ for very small $k$. Substituting Eq. (S36) in Eq. (12) in the main text, we can show that the two branches are:

$$\begin{cases} \omega_+ = -i\left(\tilde{\alpha} + \tilde{v}_0^2/\tilde{\gamma}\right)k^2 \\ \omega_- = -i\left[\left(\tilde{\eta} - \tilde{v}_0^2/\tilde{\gamma}\right)k^2 + \tilde{\gamma}\right] \end{cases}, \quad k \ll k_{EP1} \tag{S37}$$

The terms of $O(k^2)$ in $\omega_\pm$ corresponds to thermal and momentum diffusion processes. An extra term $\tilde{v}_0^2/\tilde{\gamma}$ contributes to the thermal diffusion even in the bulk, which is essentially the collective transport effect and nonnegligible for materials with strong Normal processes.

On the short wavelength side ($k \gg k_{EP2}$), the diagonal elements of $\mathcal{H}_{11}$ and $\mathcal{H}_{22}$ is on the order of $O(k^2)$ but the coupling off-diagonal part is $O(k)$, meaning the thermal-momentum coupling is negligible when probing into sufficiently small scales. As a result the two dispersion branches are simply equal to $\mathcal{H}_{11}$ or $\mathcal{H}_{22}$. A more detailed derivation below shows that the thermal diffusion and momentum dissipation switched channels. At large $k \gg k_{EP2}$, $\Omega_\theta$ can be approximated as:

$$\Omega_\theta \approx i\sqrt{\frac{(\tilde{\alpha} - \tilde{\eta})^2 k^4}{4} - \frac{\tilde{\gamma}}{2}(\tilde{\alpha} - \tilde{\eta})k^2} \approx \frac{i}{2}(\tilde{\alpha} - \tilde{\eta})k^2 - \frac{i\tilde{\gamma}}{2} \tag{S38}$$

Therefore, the dispersion of the two branches reduce to diagonal elements of $\mathcal{H}$ as follows:

$$\begin{cases} \omega_+ = -i\left(\tilde{\eta}k^2 + \tilde{\gamma}\right) \\ \omega_- = -i\tilde{\alpha}k^2 \end{cases}, \quad k \gg k_{EP2} \tag{S39}$$

From the above equation, we can see that the $\omega_+$ branch is mainly for momentum dissipation, while $\omega_-$ is dominated by the thermal diffusion in the quasiballistic APT phase.

## Note 6: Ab initio calculations of phonon properties

To obtain the phonon properties of graphite, we performed first-principles calculations using the Quantum Espresso package [36]. Harmonic force constants are calculated using the density functional perturbation theory (DFPT) using the Perdew-Zunger (PZ) exchange-correlation functionals[37] with a cutoff energy of 80 Rydberg and a 10×10×4 supercell. The third-order force constants are calculated using the finite displacement method with a 5×5×4 supercell, using the

thirdorder.py script [38] with a 24×24×10 mesh in the momentum space. We use the ShengBTE package [38] to iteratively solve the mean free displacement:

$$\frac{1}{N_q}\sum_{\lambda\nu}\left[\left(\omega_\mu \boldsymbol{F}_\mu + \omega_\lambda \boldsymbol{F}_\lambda - \omega_\nu \boldsymbol{F}_\nu\right)\Gamma^{+}_{\mu\lambda\nu} + \frac{1}{2}\left(\omega_\mu \boldsymbol{F}_\mu - \omega_\lambda \boldsymbol{F}_\lambda - \omega_\nu \boldsymbol{F}_\nu\right)\Gamma^{-}_{\mu\lambda\nu}\right] + \sum_{\lambda}\Gamma_{\mu\lambda}\left(\omega_\mu \boldsymbol{F}_\mu - \omega_\lambda \boldsymbol{F}_\lambda\right) = \omega_\mu \boldsymbol{v}_\mu \tag{S40}$$

where $\mu, \lambda, \nu$ are phonon mode indices, $\omega$ denotes the phonon frequency, $N_q$ is the total number of mesh points in the Brillouin zone, $\Gamma^{+}_{\mu\lambda\nu}$ is the transition rate for the absorption three-phonon process $\mu + \lambda \to \nu$, $\Gamma^{-}_{\mu\lambda\nu}$ denotes the transition rate for the emission three-phonon process $\mu \to \lambda + \nu$, and $\Gamma_{\mu\lambda}$ denotes the elastic isotope scattering rates calculated using Tamura's perturbation theory [39]. After obtaining the MFD, the effective relaxation times are calculated as $\tau_\mu = \boldsymbol{v}_\mu \cdot \boldsymbol{F}_\mu / \boldsymbol{v}_\mu^2$ [40]. The R-scattering rate is the summation of Umklapp scattering, isotope scattering, and boundary scattering rates: $\left(\tau_\mu^R\right)^{-1} = \left(\tau_\mu^U\right)^{-1} + \left(\tau_\mu^{iso}\right)^{-1} + \left(\tau_\mu^b\right)^{-1}$. The isotope scattering rate is calculated at natural abundance. The boundary scattering is calculated as $\left(\tau_\mu^b\right)^{-1} = v_\mu / D_{grain}$ with $D_{grain}$ the mean grain size set as 20 μm throughout this work [5,10]. The scattering rates in Fig. S1 is used to calculate second sound in the maintext.

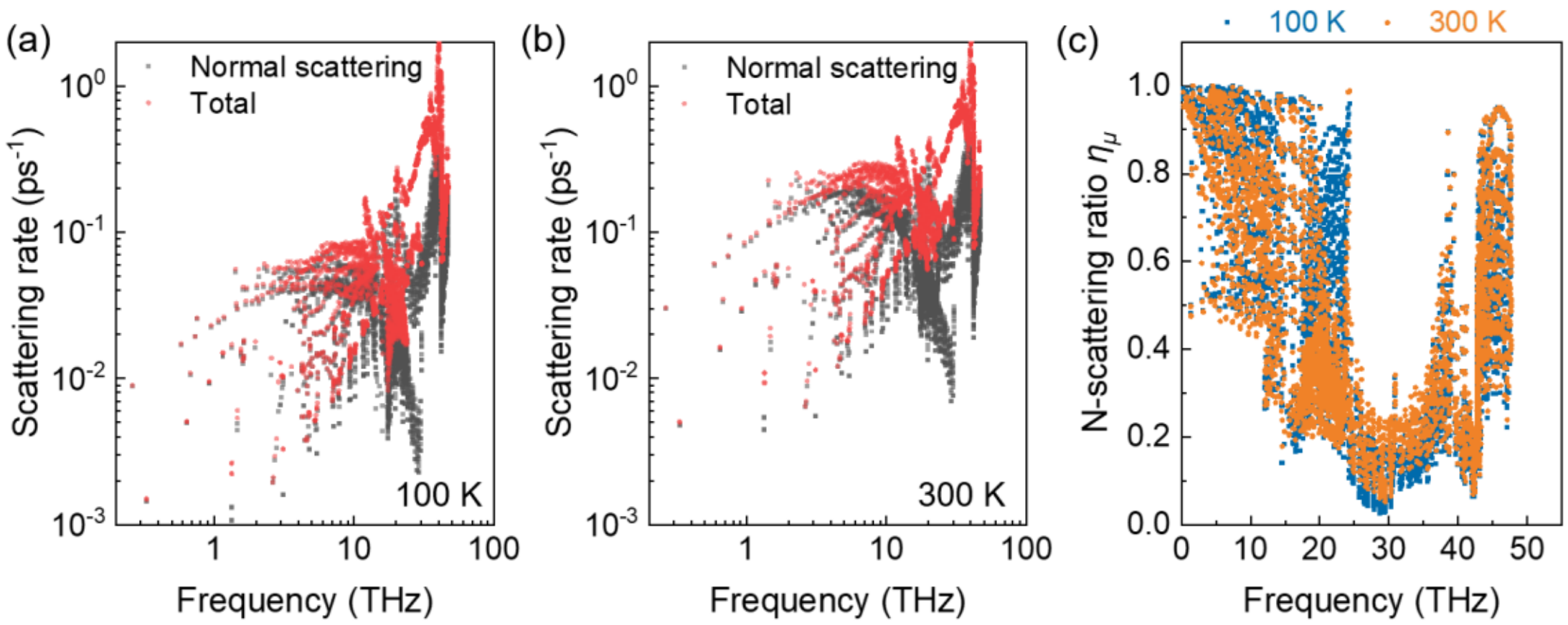

FIG. S1 *Ab initio* phonon scattering rates at (a) 100 K and (b) 300 K. (c) The relative strength of N-scattering, quantified by the ratio $\eta_\mu = \tau_\mu / \tau_\mu^N$.